\documentclass[aps,prl,twocolumn,groupedaddress,nofootinbib]{revtex4-2}

\usepackage{amsmath}
\usepackage{amssymb}
\usepackage{graphicx}
\usepackage[colorlinks=true, allcolors=blue]{hyperref}
\usepackage[dvipsnames]{xcolor}
\usepackage[normalem]{ulem}

\begin{document}
	
	\title{The cosmic trimmer:\\ Black-hole hair in scalar-Gauss-Bonnet gravity is altered by cosmology}

    \author{Eugeny Babichev}
	\email[]{babichev@ijclab.in2p3.fr}
	\affiliation{Université Paris-Saclay, CNRS/IN2P3, IJCLab, 91405, Orsay, France}

    \author{Ignacy Sawicki}
	\email[]{sawicki@fzu.cz}
	\affiliation{CEICO, Institute of Physics of the Czech Academy of Sciences, Na Slovance 1999/2, 182 00, Prague 8, Czechia.}
 
	\author{Leonardo G. Trombetta}
	\email[]{trombetta@fzu.cz}
	\affiliation{CEICO, Institute of Physics of the Czech Academy of Sciences, Na Slovance 1999/2, 182 00, Prague 8, Czechia.}
	
	\date{\today}
	
	\begin{abstract}
   	Static black holes in general relativity modified by a linear scalar coupling to the Gauss-Bonnet invariant always carry hair. We show that the same mechanism that creates the hair makes it incompatible with a cosmological horizon. Other scalar-tensor models do not have this problem when time dependence of the scalar provides a natural matching to cosmology. Scalar-Gauss-Bonnet gravity is particularly rigid, and such a scenario does not help. An extra operator makes the theory behave like the other models, and the cosmological horizon can be accommodated.  The hair, however, is drastically altered.
	\end{abstract}
	
	\maketitle

\section{Introduction}

The detection of gravitational waves (GWs) emitted from the final stages of the inspiral of binary systems of compact objects~\cite{LIGOScientific:2016aoc,LIGOScientific:2017vwq} has opened a new and finally direct avenue for testing the validity of general relativity (GR) as the theory of gravity. These tests, however, require alternative models of gravity which could be used as a laboratory for the prediction of realistic deviations from GR phenomenology. The existence of hair attached to scalar charges carried by compact objects in scalar-tensor theories of gravity provides just such a opportunity. This is the existence of alternative solutions for starlike objects in addition to or instead of the Schwarzschild or Kerr cases possible for the star's exterior in GR. These hairy (``scalarized'') solutions were discovered in the case of conformally coupled theories of gravity, where they are only possible for neutron stars but not black holes (BHs) \cite{Bocharova:1970skc,Damour:1993hw} (also see review \cite{Doneva:2022ewd}). Despite the no-hair theorems (see, e.g., \cite{Bekenstein:1995un,Hui:2012qt}), even black holes can carry hair under certain circumstances -- for example, when the theory of gravity includes interaction terms between the scalar and higher-order curvature corrections, such as the Gauss-Bonnet term (GB), in principle expected not in the least from effective field theory (EFT) arguments. This alternative phase usually exists only for ranges of BH parameters and appears as a (tachyonic) instability of the GR solution, and it can lead to modifications of observable properties of the black holes, such as changes in the spectrum of quasinormal oscillations or modifications of the BH exterior.

However, the embedding of such setups in cosmological backgrounds presents a challenge. Typically, the parameters which allow for hair in high-curvature environments while permitting a return to GR far from the objects also cause the equivalent tachyonic instability to appear in the early Universe. The result is a catastrophic modification of cosmological history compared to the one predicted from GR \cite{Anson:2019uto}. This is incompatible with observations. Modifications of the theory can push the instability into the past, but not remove it \cite{Antoniou:2020nax}. Perhaps uniquely, the linear coupling to the GB term is not known to suffer from this early-Universe instability. This theory is quite particular, in that it is the finiteness of physical observables on the BH horizon rather than the tachyonic instability that \emph{requires} that BHs carry a charge and the associated hair. This has led to a significant body of work studying this model and its new physics, including the gravitational wave signature \cite{Yagi:2011xp}. For a review, see~\cite{Babichev:2016rlq, Lehebel:2018zga, Herdeiro:2015waa} and references within.

Notwithstanding cosmological evolution, attempts have been made to find numerically a static BH solution in scalar-GB gravity in a de Sitter background. They have not been successful \cite{Bakopoulos:2018nui} for so-far unclear reasons. 

In this letter, we uncover a new type of inconsistency occurring in the presence of a second horizon, that was not understood previously. We discuss the case of a linear scalar coupling to GB (sGB) in the context of a black hole embedded within a cosmology under the assumptions of static spacetime. This model does not suffer from the tachyonic instability. We nonetheless demonstrate that the same finiteness condition that creates the BH hair makes the cosmological/de Sitter horizon incompatible with it. This \emph{two-horizon problem} is the reason why such solutions cannot be found. As a result of the shift symmetry, we can introduce a linear time dependence for the scalar without affecting observables, which has the benefit of being a natural setup in the cosmological embedding. However, this does not solve the two-horizon problem for this theory. We conclude that a nonevolving solution in this setup cannot be found.

We show an explicit example of a shift-symmetric theory where this stationary setup does work. We 
propose that the problem in linear sGB can be solved by a small extension of the model with additional shift-symmetric operators together with the cosmologically induced time dependence. In this case, the horizons no longer provide any constraint on the value of the scalar charge. We expect that this new setup would introduce substantial changes to the dynamics of black holes when compared to the static solutions in linear sGB. 

\section{Static hair in scalar-Gauss-Bonnet theory}

It is instructive to first revisit the usual static hairy black-hole solutions in the standard linear sGB theory,
\begin{equation} \label{action-sGB}
    S =M_P^2 \int d^4 \! x \sqrt{-g} \left[ \frac{R}{2} - \frac{1}{2} (\partial \phi)^2 +  \alpha \phi \, \mathcal{G} \right]\,,
\end{equation}
where $\mathcal{G} = R_{\mu\nu\rho\sigma} R^{\mu\nu\rho\sigma} - 4 R_{\mu\nu} R^{\mu\nu} + R^2$ is the Gauss-Bonnet invariant, $\phi$ is dimensionless, and $M_P$ is the Planck mass. The corresponding scalar field equation is
\begin{eqnarray}
    \square \phi + \alpha \mathcal{G} &=& 0 \,. \label{scalar-eq-sGB}
\end{eqnarray}
The size of the backreaction caused by the sGB term on the background geometry is dictated by the ratio $\alpha/L^2$, where $L$ is a characteristic curvature scale of the spacetime \cite{Sotiriou:2014pfa}. Notice that, since sGB is a shift-symmetric Horndeski theory~\cite{Kobayashi:2011nu}, the scalar field equation \eqref{scalar-eq-sGB} can be written as
\begin{eqnarray} \label{shift-symm-EOM}
    \nabla_\mu J^\mu = 0 \,,
\end{eqnarray}
where the contribution of the sGB term to the current $J^\mu$, while not unique in general \cite{Yale:2011usf} (see also \cite{Creminelli:2020lxn}), it can, for example, be obtained from the quintic Horndeski current with $G_5 \sim \log(X)$.

For a static, spherically symmetric background metric and scalar of the form
\begin{eqnarray} \label{sph-static-ansatz}
    ds^2 = - h(r) dt^2 + \frac{dr^2}{f(r)} + r^2 d\Omega^2\,, \qquad \phi = \phi(r) \,,
\end{eqnarray}
the scalar field equation can easily be integrated to give
\begin{eqnarray} \label{sGB-Jr-static}
    J_r = -\phi'+\frac{4\alpha h'}{r^2 h}(1-f)=\frac{C}{r^2 \sqrt{h f}} \,,
\end{eqnarray}
where $C$ is an integration constant to be determined. In the static asymptotically flat limit, $f,h\rightarrow 1$, and Eq.~\eqref{shift-symm-EOM} implies that $\phi\sim C/r$, with $C$ called the scalar charge of the black hole. In order to fix the value of this constant, the standard procedure is to demand the finiteness of $\phi'$ and $\phi''$ at the black-hole horizon, as proxies for observable quantities. Alternatively, a more covariant and unambiguous way to achieve the same is to require that the kinetic term \mbox{$X = - \frac{1}{2} g^{\mu\nu}\partial_{\mu} \phi \partial_{\nu} \phi$} remain finite, as it appears, for instance, in the stress-energy tensor. Near a horizon, this goes like
\begin{eqnarray} \label{static-X-near-H}
    X = - \frac{f \phi'^2}{2} \simeq - \frac{1}{2h r_h^4} \left( C - 4\alpha \sqrt{\frac{f}{h}} h' \right)^2 \,,
\end{eqnarray}
and it remains finite only if 
\begin{eqnarray} \label{static-charge}
    C = 4\alpha \sqrt{\frac{f}{h}} h' = 8 \alpha \, \kappa_\mathrm{BH} \,,
\end{eqnarray}
where we have identified the finite quantity \mbox{$\kappa = \frac{1}{2} \lim_{r \to r_h}{\sqrt{\frac{f}{h}} h'}$} as the black hole's surface gravity, which in Schwarzschild is $\kappa_\text{BH}=(2r_s)^{-1}$.

The black holes in this theory therefore carry a secondary scalar charge -- that is, it is not a free parameter but is instead fixed in terms of the black hole's surface gravity, as given by Eq.~\eqref{static-charge}. Importantly, it is the presence of a horizon that is responsible for the charge, while horizonless sources such as neutron stars can only support a much-faster-decaying scalar profile \cite{Yagi:2015oca}.

Perturbative corrections to the metric have been studied, even including slow rotation \cite{Ayzenberg:2014aka}, while fully self-consistent numerical solutions in have been studied in Ref.~\cite{Sotiriou:2014pfa}. The observed absence of dephasing on the gravitational-wave waveform during the inspiral phase of binary black-hole mergers has established the most stringent bound on the sGB coupling so far to be $\alpha \lesssim (1.2 \, \text{km})^2$ \cite{Lyu:2022gdr}. This justifies a perturbative treatment for $\alpha \ll r_s^2$ for astrophysical black holes. Further studies have also constrained the validity of the EFT [Eq.~\eqref{action-sGB}], which must have a UV cutoff $\Lambda_{UV} < 1/\sqrt{\alpha}$ in order to avoid resolvable superluminal propagation \cite{Serra:2022pzl}. 

Now, let us extend the model to include a positive cosmological constant $\Lambda>0$, producing a still static but expanding spacetime with Hubble constant $H$. For observationally allowed values of $\alpha$, $H$ is completely driven by $\Lambda$, and the correction to the speed of gravitational waves is completely negligible. Whenever the backreaction is small, we would expect the Schwarzschild-de Sitter GR solution for the spacetime to be recovered -- i.e.,
\begin{eqnarray} \label{Schw-dS}
    h = f = 1 - \frac{r_s}{r} - H^2 r^2\,.
\end{eqnarray}
In this letter, we only require that the solution~\eqref{Schw-dS} be recovered asymptotically at cosmological distances, but we allow for backreaction close to the black hole. In the astrophysically relevant scenario  $Hr_s \ll 1$, the black-hole horizon remains at $r_h\simeq r_s$, but $f,h$ also have second roots and therefore a second horizon -- the standard Hubble horizon of cosmology located at $r_h\simeq H^{-1}$.  For a scalar-field solution valid in the whole observable region, we  also have to enforce the finiteness of $X$ on the new cosmological horizon. Equation~\eqref{static-charge} then requires that $C$ simultaneously take different values at each of the horizons, while being a constant, and therefore it is impossible to satisfy. This is the origin of the nonconstruction of the black-hole solution in the presence of a positive $\Lambda$ in Ref.~\cite{Bakopoulos:2018nui}. In the next section, we will demonstrate that in cosmology, the scalar field cannot remain static, and therefore the time dependence must be taken into account when attempting to solve the two-horizon problem.

\section{Time-dependent hair in cosmologically embedded black holes}
Cosmology suggests a generalization of this setup which might resolve the problem: it is natural to consider a time-dependent scalar field.  We take the simplest ansatz which  contains both a time dependence and the spherically symmetric metric required by the local source,
\begin{equation}
\label{ansatz}
    \phi = q\,t + \varphi(r)\,,
\end{equation}
 where $q$ is a new parameter which might allow both the horizons to be compatible with the solution. Owing to the invariance of the action \eqref{action-sGB} under shifts of $\phi$, such linear time dependence does not give rise to time dependence of any observable quantity. Thus, it is compatible with static solutions in principle, as long as the effect of accretion can be neglected. As for the static case, it is again possible to integrate the scalar field equation Eq.~\eqref{shift-symm-EOM} once, giving
\begin{eqnarray} \label{sGB-Jr-rolling}
    J_r = -\varphi'+\frac{4\alpha h'}{r^2 h}(1-f)=\frac{C}{r^2 \sqrt{h f}}\,,
\end{eqnarray}
where again $C$ is a constant of integration yet to be determined. Incidentally, this is the same equation as for the static case, up to the replacement $\phi \to \varphi$. 

Following the same logic as for the static case, we will demand that the scalar quantity $X$ be finite at both horizons. With the ansatz above, the kinetic term now is
\begin{eqnarray}
    X = \frac{1}{2} \left( \frac{q^2}{h} - f \varphi'^2 \right)\,,
\end{eqnarray}
which implies that, unlike in  the static case, $\varphi$ can no longer be finite itself if $X(r_h) = X_h$ is not to diverge. Since no observable quantity depends on  $\phi$,  while $X$ is explicitly contained in at least the energy-momentum tensor, we demand only the finiteness of $X_h$ and parametrize the $r$-dependent part of the scalar near any horizon as
\begin{eqnarray}
\label{phiprime}
    \varphi' &=& \epsilon \frac{q}{\sqrt{h f}} + \psi'(r)\,,
\end{eqnarray}
with $\epsilon(r_h) = \pm 1$, and where $\psi(r_h)$ is now manifestly finite and related to $X_h$ as 
\begin{eqnarray}
    X_h = - \epsilon_h q \sqrt{\frac{f}{h}} \, \psi'(r_h) \,.
\end{eqnarray}
Then in this near-horizon limit the scalar field equation~\eqref{sGB-Jr-rolling} multiplied by $\sqrt{h f}$ reads 
\begin{eqnarray} \label{horizon-sGB}
    \frac{C}{r_h^2} = \sqrt{\frac{h}{f}} J^r|_h &=& - \epsilon_h q + 4 \alpha \sqrt{\frac{f}{h}} \frac{h'}{r_h^2} + O(h, f) \,. 
\end{eqnarray}
Notice that at this order, there is no dependence on $X_h$, or equivalently $\psi'_h$. It can now easily be checked that the scalar quantity $J^2$ remains finite at the horizons for any value of $C$.\footnote{This follows, provided that $\frac{f'}{f} - \frac{h'}{h}$ is regular, which is in turn necessary to avoid a curvature singularity. Then \mbox{$J^2 \sim 1/q$}, which is finite for $q \neq 0$.} This is contrary to the static case, where the divergence was unavoidable already at the BH horizon~\cite{Babichev:2016rlq}. 

Now let us, in turn, examine each horizon and see which conditions arise on $C$. The condition on the black-hole horizon gives
\begin{eqnarray} \label{BH-horizon-sGB}
     C &=& r_s^2 \left( - \epsilon_h q + \frac{8 \alpha}{r_s^2} \kappa_\mathrm{BH} \right) \,.
\end{eqnarray}
Choosing $C$ requires that we pick some $q$ -- this is the extra freedom we have introduced. We can do so by examining the cosmological horizon: Eq.~\eqref{horizon-sGB} naively gives
\begin{eqnarray} \label{cosmo-horizon-sGB}
    q &=& H^2 C + 8 \alpha H^3 + O(H r_s) \,,
\end{eqnarray}
where it is assumed that the geometry approaches de Sitter. Seemingly, we have found our new solution compatible with both horizons. The question now is how this connects to the cosmological solution. 

Let us discuss the theory [Eq.~\eqref{action-sGB}] in a cosmological setting in the absence of a local source. For simplicity, we consider de Sitter space,
\begin{eqnarray}
    ds^2 = - d\tau^2 + e^{2H\tau} \left( d\rho^2 + \rho^2 d\Omega^2 \right) \,,
\end{eqnarray}
where the scalar-field equation takes the simple form
\begin{eqnarray}\label{cosmo-full-EOM-SGB}
    \ddot{\phi} + 3H \dot{\phi} = 24 \alpha H^4 \,,
\end{eqnarray}
with $\dot{} = \partial_{\tau}$. This equation admits an attractor ($\ddot{\phi} = 0$) solution
\begin{eqnarray} \label{cosmo-EOM-sGB}
    q_c \equiv \dot{\phi} = 8  \alpha H^3 \,.
\end{eqnarray}
Thus, the cosmology imposes that the scalar field always asymptotes to a rolling stationary solution
\begin{eqnarray} \label{cosmo-solution}
    \phi_c = q_c\, \tau\,.
\end{eqnarray}
Translating the cosmological coordinates to the static ones, this implies that in the $r\rightarrow\infty$ limit, the cosmological $\phi$ tends toward
\begin{eqnarray}\label{qc-in-static}
    \phi = q_c \, t + \int \mathrm{d}r \varphi '\,,\quad \varphi' \simeq \frac{q_c}{Hr} \,.
\end{eqnarray} 
The only consistent possibility when embedding a BH in cosmology is that the solution sufficiently far away from the BH becomes homogeneous. We thus need to match the scalar hair to this cosmological boundary condition [Eq.~\eqref{cosmo-solution}] sufficiently far away. This requires an $r$ dependence as in Eq.~\eqref{qc-in-static}, which is satisfied by the solution of Eq.~\eqref{sGB-Jr-rolling} for $r\rightarrow\infty$, while the ``local'' time dependence $q$ only contributes decaying terms in observable quantities,\footnote{We thank Dra\v{z}en Glavan for useful comments about this point.} e.g., $X - X_c \simeq e^{-2H\tau} (q-q_c)^2/2(H \rho)^2$. A similar behavior was observed for Brans-Dicke theory \cite{Glavan:2019yfc}. However, in the case of sGB, the finiteness condition fixes $q$ as
\begin{eqnarray} \label{eq:q-vs-qc}
    \frac{q}{q_c} = \frac{1}{2(H r_s)} \frac{\kappa_\mathrm{BH}}{\kappa_s} + 1 + O(H r_s) \gg 1 .
\end{eqnarray}  
This is vastly different from the cosmological value $q_c$, implying either (a) a very slow approach to homogeneity only at distances $H \rho \sim 1/(H r_s) \gg 1$, or (b) a backreaction so strong as to completely screen the gravitational effect of the black hole -- i.e., $\kappa_\textrm{BH} \lesssim (Hr_s) \kappa_s \ll \kappa_s$. Moreover, this tension is larger for small black holes. As a matter of principle, the local solutions should not affect the cosmology strongly, but rather the cosmology should represent the aggregated effect of the averaged localized sources -- this principle is violated by the stationary solution in sGB.

This is the main result of this letter: There is an impossibility to construct black-hole solutions in the most simple sGB theory [Eq.~\eqref{action-sGB}] with a physically acceptable approach to the homogeneous cosmological asymptotics, when the metric is static and the scalar field is time dependent at most in a manner which keeps observables static. The problem stems from the $r_s^{-1}$ scaling of the required values of $C$ and $q$ to keep scalar quantities ($X$) finite at both the black-hole and cosmological horizons, making it incompatible with having both the usual notion of a black hole and cosmological homogeneity at reasonable separation.

One may think that the two-horizon problem could be fixed without much change to the expected physics by breaking the shift symmetry and adding a potential $V(\phi)$ to the scalar field. In this case, the scalar can be asymptotically static at the minimum of $V_{,\phi}+\alpha \mathcal{G}$. Picking $V(\phi)$ such that the mass is of inverse galactic scales would then screen the hair at cosmological distances while giving astrophysics as expected in the static shift-symmetric case. This is true -- this modification would allow us to use the static solution in the completely static universe. However, during the approach to the asymptotic future de Sitter state, the minimum of the potential evolves and the scalar must roll slowly, giving a small $\dot\phi =q$. This would then be enough to put us in the scenario in which the finiteness of $X$ on the horizon would imply that $\phi$ diverges. Since $\phi$ is now a physical observable, this is inconsistent. It remains to be seen whether any such non-shift-symmetric resolution of the problem would not alter the nature of the standard static solution. A similar conclusion can be reached for a mass term provided by a coupling to the Ricci scalar $\phi^2R$.

Next, we investigate an alternative approach to solving the problem that maintains shift symmetry.

\section{The good example: Cubic Galileon}

A fair question is whether this type of problem appears in other scalar-tensor theories. This turns out not to be the case in general. As an example of a model which is free of this particular issue, we now consider the cubic Galileon model discussed in Ref.~\cite{Babichev:2012re}, with the action
\begin{equation} \label{action-cubic}
    S = M_P^2\int d^4 \! x \sqrt{-g} \left[ \frac{R}{2} + \frac{1}{2} (\partial \phi)^2 + \frac{(\partial \phi)^2}{2\Lambda^2} \square \phi \right] \,.
\end{equation}
Notice that the kinetic term has a ghostly sign, which leads to self-accelerated cosmological solutions \cite{Deffayet:2010qz}. The homogeneous solution for the scalar of the form \eqref{cosmo-solution} asymptotes toward
\begin{eqnarray} \label{cosmo-q-KGB}
     q_c = - \frac{\Lambda^2}{3H}\,.
\end{eqnarray}
Now, turning on the mass of the black hole, we consider as previously the ansatz~(\ref{ansatz}) for the Schwarzschild-de Sitter background metric. The analogue of Eq.~\eqref{sGB-Jr-rolling} in this case is
\begin{equation}
\label{cubic}
    J_r = \varphi'+\frac{1}{\Lambda^2}\left(\frac{X h'}{h}-\frac{2f\varphi'^2}{r}\right)=\frac{C}{r^2 \sqrt{h f}}\,.
\end{equation}
The above expression cannot be satisfied at all unless $X$ is finite, implying a solution of the form~(\ref{phiprime}). This means that the value of $X$ is \emph{necessarily} finite at a horizon for any $C$. This is in contrast to the case of sGB theory, where we have to fix $C$ and $q$ to avoid the divergence of $X_h$. The value of $X_h$ at a horizon is found from Eq.~\eqref{cubic} by taking the limit $r\to r_h$, 
\begin{eqnarray} \label{horizon-KGB}
    \frac{C}{r_h^2} = 
    \epsilon_h q + \frac{1}{\Lambda^2} \sqrt{\frac{f}{h}} \left( X_h h' - \frac{2q^2}{r_h} \right) + O(h,f) \,.
\end{eqnarray}
This imposes no condition on either $C$, $q$, or $X_h$, which may or may not be subject to other constraints away from the horizons.

\section{Extra operators alleviate the problem}

The fact that the example of the cubic Galileon works well, in contrast to the sGB case, can be traced to the form  of the current~(\ref{cubic}).
The explicit presence of $X$ in this equation forces $X_h$ to be finite. In this respect, the sGB operator is rather unique among the Horndeski operators in not providing an explicit $X$ term in the current \eqref{sGB-Jr-rolling}.  
The natural question is then whether the presence of any of these additional operators may mitigate the problem faced by the vanilla sGB theory [Eq.~\eqref{action-sGB}]. 

To have an explicit example, it is not difficult to check that when combining the sGB term with the cubic Galileon term, the solution for the scalar field remains of the form~(\ref{phiprime}), thus providing a finite $X_h$ at both horizons. We expect similar results when we combine sGB with other Horndeski terms that contain a function of $X$. Indeed, assuming the behavior in Eq.~\eqref{phiprime}, for the sGB theory with an extra shift-symmetric term, the equation of motion for the scalar at a horizon becomes [cf.\ Eq.~(\ref{horizon-KGB})],
\begin{eqnarray} \label{extra-ops-C}
    \frac{C}{r_h^2} = - \epsilon_h q + F\left(q, X_h\right) + 4 \alpha \sqrt{\frac{f}{h}} \frac{h'}{r_h^2} + O(h,f),
\end{eqnarray}
where $F$ depends on the four Horndeski functions $G_i$. As one can see, for arbitrary values of $C$ and $q$, the solution $\varphi'$ exists with finite $X$ at horizons and our modified sGB behaves just like the cubic Galileon. It must be stressed that our argument does not establish for which particular values of $C$ solutions regular in the whole intermediate spacetime actually exist. Determining when this construction is viable is beyond the scope of this letter.

One may still think that picking $C$ as in the static case~\eqref{static-charge} and $q=0$ would give physics similar to the static case. This is not true. Notwithstanding whether such a solution would be regular, the background for $\phi$ would now be very different, as $F$ would have to make a significant contribution to solve the two-horizon problem and, as a consequence, the perturbations would have very different dynamics. GW observables such as radiated power and the quasi-normal-mode spectrum would be strongly affected.

\section{Discussion \& Conclusions}

Let us take a different look at what has happened here: it is instructive to discuss shift-symmetric theories like sGB in terms of the dynamics of the corresponding shift charges -- the equation of motion for the scalar \eqref{shift-symm-EOM} is a conservation equation for the shift-charge current, $J^\mu$~\cite{Pujolas:2011he}, where $J^0$ is the shift-charge density and  $J^i$ is the spatial current giving the motion of these charges. The static sGB solution then corresponds to a static radial current \eqref{sGB-Jr-static} with \emph{flux} $C$ across any surface enclosing the horizon.\footnote{We stress that, in the shift-charge picture, $C$ is a flux of conserved shift charges the density of which is given by $J^0$, not a scalar charge as it is usually referred to.} For this static configuration to be maintained, the charges must leave the spacetime at the rate they enter, and they are able to do so at spatial infinity. Keeping $X$ finite on a horizon means that the horizon must produce or absorb  a particular flux $C$ of charges given by Eq.~\eqref{static-charge}. Introducing a second horizon, e.g.,\ cosmology, gives a different flux requirement. The two are in general incompatible, so the charges will begin to amass in the intermediate spacetime between the horizons breaking the staticity assumption of the solution. 

Cosmology introduces time dependence to the scalar, with an attractor solution $q=q_c$ [Eq.~\eqref{cosmo-EOM-sGB}]. This is also a vacuum solution with $J^\mu=0$ to which any black hole would have to eventually match. This requires stationary time-dependent hair [Eq.~\eqref{BH-horizon-sGB}]. However, the requirement of finiteness on both horizons requires a $q$ grossly different from $q_c$, and therefore a shift-charge cloud extending as far as  $r\sim1/H^2r_s$, much beyond the cosmological horizon. 

We have not found any reasonable stationary solutions in sGB under our assumptions, but it may well be that the evolution away from the static solution is slow, and therefore one might be able to use the $q\simeq q_c$ solution, with $C$ given by \eqref{BH-horizon-sGB} for sufficiently long to study, e.g.,\ binary inspirals. Even so, the static solution is not a good approximation, since the cosmological background provides a medium which may affect the dynamics.

Alternatively, the introduction of (an) extra Horndeski terms resolves the inconsistency of the requirements put on the solution. The presence of $X$ in Eq.~\eqref{extra-ops-C} means the theory is less rigid and allows for any choice of flux $C$ across any horizon, even in the presence of the \mbox{$q \simeq q_c$} compatible with the cosmological vacuum needed to be reached at horizon scales. This is, of course, not sufficient to guarantee that such a solution exists -- it only provides the boundary conditions on the two horizons, which must then be connected through a non-singular source-free solution. This underlines how spectacularly the \emph{static} sGB solution in cosmology fails -- it cannot even support the required boundary conditions. 

The price of this well-behaved alternative is that the hair which is necessary once a horizon appears in sGB Eq.~\eqref{static-charge} is no longer the same upon the introduction of any generic additional Horndeski operator. In any such good solution, the background that this hair gives for the fluctuations will be drastically different from the static sGB solution, and we expect that it will strongly affect the predictions of any physical process. In particular, predictions for GW signatures would need to be reassessed in light of our findings. We leave the construction of a viable configuration and the study of its properties to a separate work.

\begin{acknowledgments}
\section{Acknowledgments}
The authors would like to thank Christos Charmousis, Gilles Esposito-Farèse, Dra\v{z}en Glavan, and Nicolas Lecoeur for helpful discussions. E.B.\ acknowledges the support of ANR grant StronG (ANR-22-CE31-0015-01). The work of L.G.T.\ was supported by the European Union (Grant No.\ 101063210).  I.S.\ is supported by the Czech Science Foundation (GAČR) project PreCOG (Grant No.\ 24-10780S). E.B.\ and L.G.T.\ would like to thank the hospitality of the Centro de Ciencias de Benasque Pedro Pascual during the initial stages of this work.
\end{acknowledgments}

\bibliography{refs}

\end{document}